\documentclass[traditabstract]{aa}
\usepackage{graphicx}
\usepackage{txfonts}
\usepackage{multirow}
\usepackage{longtable}
\usepackage[authoryear]{natbib}
\bibpunct{(}{)}{;}{a}{}{,}
\begin{document}
\newcommand{\kms}{km\,s$^{-1}$}

  \title{Aluminium oxide\\
        in the optical spectrum of VY Canis Majoris}

  \author{ T. Kami\'{n}ski\inst{\ref{inst1}},
                M. R. Schmidt\inst{\ref{inst2}}, and 
                K. M. Menten\inst{\ref{inst1}} } 

  \offprints{T. Kami\'{n}ski}

  \institute{\centering 
                    Max-Planck-Institut f\"ur Radioastronomie, Auf dem H\"ugel 69, 53121 Bonn, Germany,\\
                    \label{inst1}
                    \and 
                    Nicolaus Copernicus Astronomical Center, Polish Academy of Sciences, Rabia\'nska 8, 87-100 Toru\'n, Poland,\\
                    \label{inst2}
                   \email{kaminski@mpifr.de, schmidt@ncac.torun.pl, menten@mpifr.de} %
            }
	    
  \date{Received; accepted}

\abstract{We report the first identification of the optical bands of the $B\,^2\Sigma^+$--$X\,^2\Sigma^+$ system of AlO in the red supergiant VY\,CMa. In addition to TiO, VO, ScO, and YO, which were recognized in the optical spectrum of the star long time ago, AlO is another refractory molecule which displays strong emission bands in this peculiar star. Simulating the bands of AlO, we derive a rotational temperature of the circumstellar gas of $T_{\rm rot}$=700\,K. By resolving individual rotational components of the bands, we derive the kinematical characteristics of the gas, finding that the emission is centred at the stellar radial velocity and its intrinsic width is 13.5\,\kms\ (full width at half maximum). It is the narrowest emission among all (thermal) features observed in VY\,CMa so far. The temperature and line widths suggest that the emission arises in gas located within $\sim$20 stellar radii, where the outflow is still being accelerated. This result contradicts equilibrium-chemistry models which predict substantial AlO abundances only to within a few stellar radii. We argue that non-equilibrium models involving propagation of shocks are needed to explain the observations.} 
\authorrunning{Kami\'nski, Schmidt, \& Menten}
\titlerunning{Optical AlO emission in VY\,CMa}
\maketitle

\section{Introduction} \label{intro}
VY Canis Majoris (VY\,CMa) is a peculiar red supergiant with a very high luminosity of 3$\cdot10^5$\,L$_{\sun}$, an effective temperature,  $T_{\rm eff}$, of 3200--3500\,K, and an enormous size which may be up to 3000\,R$_{\sun}$ \citep{wittkowski}. It is thought to be an evolved massive star with initial mass of 25$\pm$10\,M$_{\sun}$ \citep{wittkowski}.  That mass was reduced to the current  17$\pm$8\,M$_{\sun}$ \citep{wittkowski} due to mass-loss phenomena that occurred at different stages of its evolution. The current mass-loss rate of the order of $10^{-4}$\,M$_{\sun}$\,yr$^{-1}$ \citep{danchi} is probably the highest in VY\,CMa's history and resulted in an extended ($\sim$7\arcsec) reflection and emission nebula that partially obscures the object \citep{smith01,h05}.  So far unknown remains the mechanism that elevates matter high enough above the stellar surface to cool it down to temperatures at which  dust grains can condense. Once dust is formed, radiation pressure on grains can drive the powerful circumstellar outflow, but such a ``standard" scenario alone does not explain all the observed features of the VY\,CMa's outflow \citep{smith01}. It has been suggested that the highly complex structure of the nebula may be related to huge convective cells on the surface of the enormous star (e.g. \citealp{smith01}; see also \citealp{schwarzschild}).  

Due to its extreme parameters and the possibility that it will explode as a core collapse supernova, VY\,CMa has been a subject of considerable astrophysical interest.  This has recently intensified due to detections of new molecules in its cold circumstellar environment. While the first radio detections of oxides OH, H$_2$O, SiO date back to late-1960s/early-1970s  \citep{OHdetect,H2Odetect,SiOdetect}, it is the recent developments in millimeter and submillimeter astronomy that have allowed to discover the rich molecular spectrum of VY\,CMa at radio wavelengths. Line surveys (i.e. systematic observations of substantial frequency ranges) like those of \citet{ziurys} and \citet{tanen10} (see also \citealp{royer}) allowed to fully explore the radio spectrum yielding 23 different molecular species (plus isotopologues) presently known.
To date, the radio spectrum of VY\,CMa is the finest example of molecular richness in oxygen-rich gas around an evolved massive star. 

Optical observers have realised the uniqueness of the spectrum of VY\,CMa much sooner \citep{joy42,herbig1970} than radio-astronomers and it is the unusual optical spectrum that have triggered much of the attention garnered by this object for more than a half a century. In addition to strong atomic {\it emission} lines, whose presence is a rare phenomenon by itself, VY\,CMa displays strong optical emission in molecular bands. Such molecular emission so far has been identified in electronic bands of TiO, VO, ScO and YO \citep{hyland69,waller71,herbig74,waller86}. It indicates the presence of warm (a few hundred K) circumstellar material close to the star. All the oxides that are seen in emission in VY\,CMa are also those refractory molecules which are most often observed in absorption in the photospheres of cool stars. Another such refractory molecule, AlO, has never been reported in emission (nor in absorption) in the optical spectrum of VY\,CMa.

However, pure rotational emission from the AlO radical has been recently detected in VY\,CMa at millimeter wavelengths \citep{tanen_AlO}. Being  a refractory and also a relatively abundant molecule, AlO is of special astrochemical interest. It proceeds the formation of Al$_2$O$_3$ which is thought to be the first molecule to condense in an outflow of a late-type star like VY\,CMa \citep[e.g.][]{SaH}. Its observations bear the potential of gaining our understanding of dust condensation in outflows of cool stars, in particular of the formation of alumina dust. 
Due to an unresolved fine and hyperfine splitting of the pure rotation lines, the small number of detected transitions, and the poor angular resolution of the available radio data \citep{tanen_AlO}, the origin of AlO emission in VY\,CMa remains poorly understood. A potential detection of AlO at shorter wavelengths (i.e. in the infrared or in the optical regimes) can help to better establish how the molecule is formed and what its role is in dust formation.  

In an archival spectrum of VY\,CMa from 2001, we found strong emission features which undoubtedly belong to the $B\,^2\Sigma^+$--$X\,^2\Sigma^+$ system of AlO.  We describe the observations and the overall spectrum of VY\,CMa in Sects.\,\ref{obs} and \ref{spec}. We characterise the AlO emission briefly in Sect.\,\ref{spec} and compare it to our  simulations in Sect.\,\ref{analysis}, where we also give basic characteristics of the emitting gas derived in our analysis. We discuss the discovery of AlO optical emission and its implications in Sect.\,\ref{dis}.

\section{Observations and data reduction}\label{obs}
The spectrum of VY\,CMa was taken with the Ultraviolet and Visual Echelle Spectrograph (UVES) at the Very Large Telescope (Paranal Observatory, Chile) on 2001 April 12. Two exposures  of duration 2 and 10\,s were taken. The spectrum covers three spectral ranges, i.e. 4010--5220\,\AA\ (blue), 6460--8315\,\AA\ (red), and 8430--10\,300\,\AA\ (far red). The slit width of 1\arcsec\ provided a spectral resolution of about $R$=45\,000. We acquired the spectra through the ESO archive and reduced them using the UVES pipeline\footnote{http://www.eso.org/sci/software/pipelines/} and IRAF using standard procedures for echelle spectroscopy. With the slit length of 10\arcsec\ and the slit position angle of 0\degr, the nebula of VY\,CMa filled a substantial part of the slit and a careful correction for the background was applied. The stellar spectrum was extracted within an aperture with a width of 2\arcsec. With seeing of 0\farcs8\ (full-width half maximum, FWHM) during the observations, this aperture encloses most of the flux of the central source (the cut off was at 5\% of the peak) and avoids most of the extended nebula. In this study, we focus on the blue spectrum for which the wavelength calibration is reliable to within $\sim$0.4\,\kms\ (3\,rms) or better. Because no spectro-photometric standard star was observed with the spectral setup, the spectrum of VY\,CMa could not be properly calibrated in flux. To roughly put the final spectrum on a flux scale, we used the spectral energy distribution of VY\,CMa from \citet{massey}, which is based on observations five years after those with UVES.   
As we checked at the visual light curve by American Association of Variable Star Observers\footnote{http://www.aavso.org/lcg}, within uncertainties of 0.4\,mag, VY\,CMa had the same brightness on the date of the UVES observations and those of \cite{massey}.

\section{The optical spectrum of VY\,CMa}\label{spec}

The extraordinary appearance of the optical spectrum of VY\,CMa has been extensively described in the literature \citep[e.g.][]{hyland69,waller71,waller86,h05}. In particular, a spectrum obtained a year before the UVES observations, at a similar resolution, and covering the range 5175--9140\,\AA\ was described in detail by \citet{waller2001}.  We find no significant differences in both spectra in the overlapping region, and therefore here outline only briefly the main spectral features. The photospheric (absorption) spectrum is dominated by molecular bands of TiO as expected for an early to intermediate type M-type star. On top of this photospheric spectrum, VY\,CMa displays a great variety of emission features which include atomic lines and molecular bands. The atomic emission is dominated by neutral atoms with lines of \ion{Ba}{II} being the only known exception. Strongest is emission from resonance lines of \ion{K}{I}, \ion{Na}{I}, \ion{Ca}{I}, and \ion{Rb}{I} that form classical P-Cyg profiles. Many of the weaker emission features (e.g. from \ion{Ti}{I}, \ion{Cr}{I}, \ion{V}{I}, and \ion{Fe}{I}), exhibit inverse P-Cyg profiles or appear as pure emission lines. The spectrum contains also broad (non-photospheric) atomic absorption lines, which may be partially filled with emission.  

The UVES spectrum contains strong and numerous emission bands of TiO and at least one band of VO (at 7867\,\AA) which all appear unchanged when compared to the spectrum of \citet{waller2001}. Unfortunately, our spectrum does not cover the strongest bands of ScO and YO, which were present in the spectrum of \citet{waller2001}, and there is no evidence for emission of these molecules within our coverage.

\begin{figure*}
\includegraphics[angle=270,width=\textwidth]{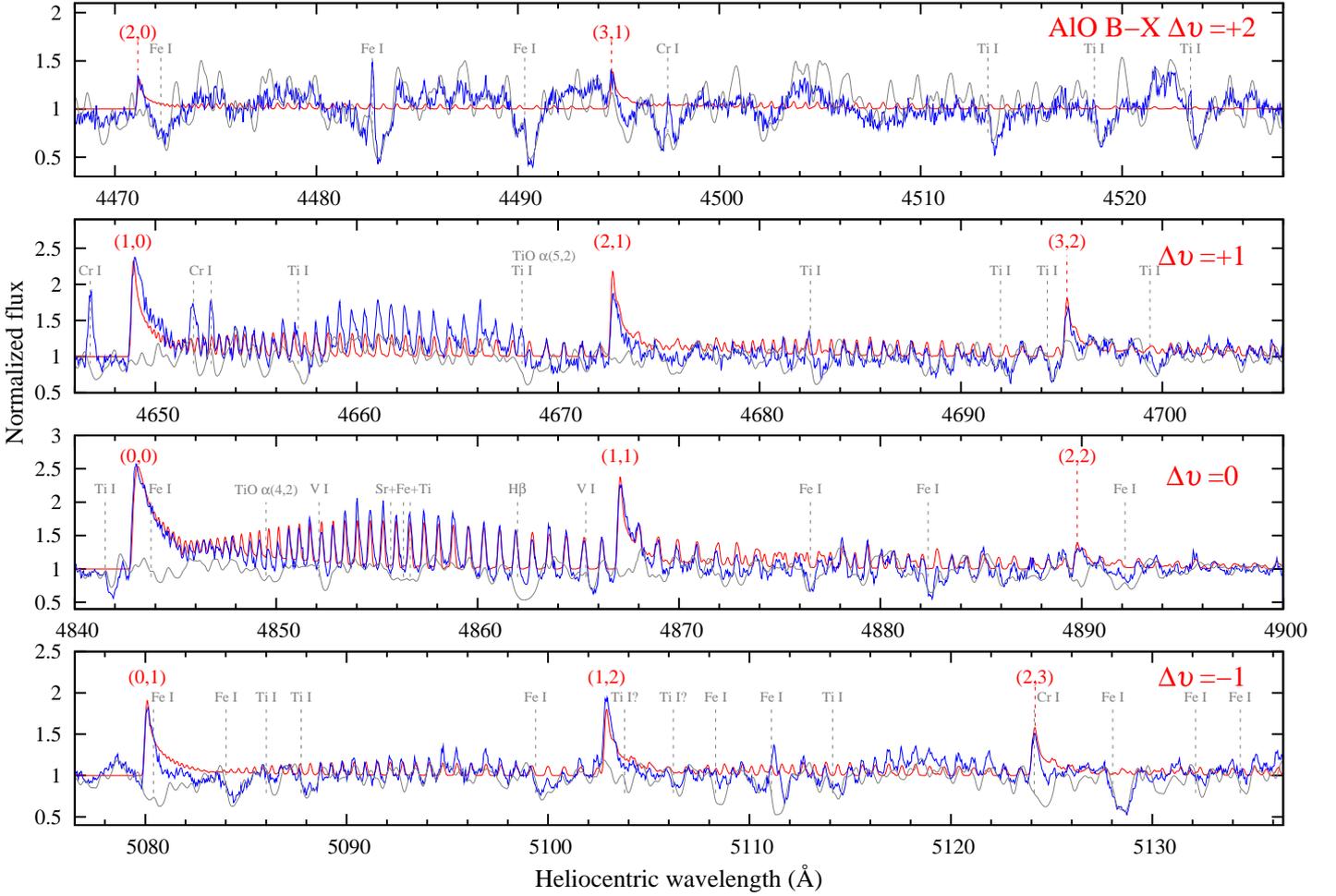}
\caption{The simulation of the $B$--$X$ system of AlO (red line) is compared to the spectrum of VY\,CMa (blue line). The simulation was obtained for $T_{\rm rot}$=700\,K and $T_{\rm vib}$=2200\,K. A rescaled spectrum of Betelgeuse is also shown (grey line) to illustrate the contribution of photospheric (and circumstellar) features that have not been included in the simulation. From top to bottom, the panels contain sections of the spectrum dominated by the $\Delta\varv$=2, 1, 0, and --1 sequences. The positions of AlO band-heads are indicated with red markers and labels (giving $\varv^{\prime\prime}$ and $\varv^{\prime}$), while some of the strongest features in the stellar spectra are indicated by grey markings.}
\label{fig-AlO-bands}
\end{figure*}

The optical spectrum of VY\,CMa short-ward of about 5100\,\AA, which is partially covered by the UVES spectrum, has not been explored by observers at high resolution. Between around 4800 and 5600\,\AA, we found strong emission features with a well recognizable `wavy' pattern indicating a molecular rotation spectrum. We have identified these features as belonging to the $B\,^2\Sigma^+$--$X\,^2\Sigma^+$ (blue--green) system of AlO. A careful inspection of the spectrum revealed the presence of the $\Delta\varv$\,=\,$-$1, 0, 1, 2 sequences. The other well known sequence $\Delta\varv$\,=\,--2 with the main head above 5335\,\AA\ is not covered by the UVES spectrum and is too weak to be seen in the spectrum of \citet{waller2001}. The UVES spectrum also does not reach the $A$--$X$ system of AlO, which is located above 1.05\,$\mu$m and we do not know any published spectrum that covers this range. In Fig.\,\ref{fig-AlO-bands}, we compare the sections of the spectrum containing the AlO emission with our best simulation of the bands (see below). The $\Delta\varv$=$+$2 sequence is hardly recognisable and essentially appears only in two band-heads, i.e. ($\varv^{\prime\prime}, \varv^{\prime}$)=(2,0) and (3,1), that look similar to atomic emission features. Each of the lower overtones with $\Delta\varv$=--1, 0, 1 exhibits three to four unambiguously recognizable heads (i.e. with $\varv^{\prime\prime}\!\!\leq$3 and $\varv^{\prime}\!\!\le$4) and the molecular emission can be traced in individual rotational components even up to $\sim$100\,\AA\ from the main head. The head of the (0,0) band, which reaches a flux level of 2.5 times the local continuum, is the most prominent emission feature in the blue part of the UVES spectrum. 

\section{Analysis of the AlO $B$--$X$ system in emission}\label{analysis}

We simulated the emission in the $B$--$X$ system of AlO using analogues procedures as in \citet{kami_v4332}. We assumed that the emission comes from a volume of homogeneous gas with no velocity gradient and that upper levels are populated by absorption of photons from a central source. The pumping to upper levels is less effective for optically thick lines, because of the reduction of flux from the central source, and this effect was taken into account allowing a derivation of gas optical thickness {\it in the radial direction}. The emitting gas is parametrised by vibrational and rotational temperatures (their meanings are explained later), column density in the direction to the star, and micro-turbulence. A given value of column density determines the optical thickness of a given transition and, as described later, it is the optical-thickness effects that allow a determination of the column density. In our approach, best fit parameters do not depend directly on the flux calibration nor the dust extinction in the nebula. In order to compare the simulation results to the observations, the simulated spectrum was smoothed with a Gaussian, what may be interpreted as a convolution with a macro-turbulence profile.  

The molecular data were taken from a compilation of sources: the list of lines was built from spectroscopic constants in \citet{saksena}; the electronic transition moment was adopted from \citet{zeno}; Franck-Condon factors were taken from \citet{coxon}; rotational line-strength factors were computed numerically following \citet{hougen} and verified by comparison to the analytical formula in \citet{kovacs}. Because Al has only one stable isotope, all the analysis was performed for the single isotopologue $^{27}$Al$^{16}$O.

Since we deal with a $^2\Sigma^+$--$^2\Sigma^+$ transition (Hund's case b), each rotational level, designated by the quantum number $N$, is split into two sublevels,  $F_1$ and $F_2$, with total angular momenta given by $J$=$N$+1/2 and $J$=$N$--1/2, respectively. This is known as spin- or $\rho$-doubling. The band is formed mainly by the doublet branches $P_1$, $P_2$ and $R_1$, $R_2$. There are also satellite branches $^PQ_{12}$ and $^RQ_{21}$, which were included in the simulations, but their intensity drops very rapidly with increasing $N$ and they have no practical meaning for our analysis. The $\rho$-doubling in the $P$ and $R$ branches is taken into account in our simulations following \citet{saksena}. Although there is some controversy in the literature on the splitting constants $\gamma$ for $\varv^{\prime\prime}>0$ \citep[see e.g.][for the most recent review]{hyper}, it does not affect our analysis. 

Figure\,\ref{fig-AlO-bands} presents the results of our best simulation. The shapes of the heads, the heads-intensity ratios, and the shape of the overall rotational profile are very well reproduced by the simulation. Some minor deviations between observed and simulated spectrum are due to an imperfect spectrum normalization and a presence of photospheric and circumstellar features, which were not taken into account in the simulation. To illustrate the shape and positions of the contaminating features, we compare the spectrum of VY\,CMa to a spectrum of another red supergiant (Betelgeuse=$\alpha$\,Ori, M1\,Ia) in Fig.\,\ref{fig-AlO-bands}. This reference spectrum was extracted form the Paranal UVES Atlas \citep{POP}. The match between the photospheric and circumstellar absorption features in both stars is very good except that  some features are slightly stronger in Betelgeuse (e.g. H$\beta$) or filled with emission components\footnote{In choosing the reference photospheric spectrum, we did not attempt to find a good match to the spectral type of VY\,CMa. The general appearance of its photospheric spectrum is much better reproduced by spectral types close to M4, e.g. that of the giant HD\,11695 from the same spectral atlas.}. For VY\,CMA, the analysed spectrum can be satisfactorily explained by the combination of a cold photosphere and the emission spectrum of AlO. We also note that from the comparison of observations to the simulation, it is apparent that the AlO emission is not affected by any underlying photospheric absorption of AlO. If VY\,CMa follows the trend observed in giants\footnote{Showing AlO emission, VY\,CMa is, however, more similar to Mira variables, in which AlO intensity does not correlate with spectral type and is highly variable \citep{keenan}.}, which display AlO absorption only at spectral types later than M4 (very weak at M5) \citep{keenan}, AlO absorption is negligible in VY\,CMa, which is of spectral type M4--M5 \citep[e.g.][]{wittkowski} or earlier \citep[$\sim$M2,][]{massey}. Because the $\Delta\varv$=0 sequence is strongest and least contaminated by the photospheric and circumstellar features, it was the primary subject of the further analysis.

By obtaining a grid of simulations with variable input parameters and comparing it to the observations, we were able to constrain some physical quantities of the AlO emitting gas. 
The rotational temperature ($T_{\rm rot}$), which corresponds to the population of rotational levels within the vibrational states, can be derived from the overall shape of the rotational contours. Particularly useful for putting constraints on $T_{\rm rot}$ is the observed position of the local maximum of the emission in the (0,0) band (outside the main bandhead), which occurs at 4654$\pm$2\,\AA\ corresponding to $N$=21$\pm$2  (Fig.\,\ref{fig-AlO-bands}). It implies $T_{\rm rot}$ between 600 and 900\,K. Additionally, for rotational temperatures higher than $\sim$800\,K, the emission components of the $R$ branch are too strong at the band origin, and below $\sim$500\,K the emission of the $P$ branches forming the tail of the band is too weak. Therefore the rotation temperature implied by the simulation of the (0,0) band is in the range  $T_{\rm rot}$=700$\pm200$\,K. However, the shape and intensity of the (1,\,0) and (1,\,2) bands is better reproduced by simulations with a higher rotational temperature between 900 and 1100\,K.

Optical-depth effects also have an influence on the shape of the rotational spectrum. Because they are particularly evident in the intensity of the band-head with respect to the rest of the band, it is possible to decouple the optical thickness effects from those of varying $T_{\rm rot}$. Our best fit resulted in optical depths of $\tau_{\rm head}$=3.0, 1.0, and 0.14 in the heads of the (0,0), (1,1), and (2,\,2) bands, respectively. These values are equivalent to setting the AlO total column density to $N_{\rm AlO}$=6$\cdot$10$^{16}$\,cm$^{-2}$. Because our simulations do not contain any realistic velocity gradient, these values are upper limits on the actual optical depths and the column density if the emission arises in an outflow.  The shape of the rotational contour is also somewhat influenced by the assumed value of micro- and macro-turbulence, but we kept those parameters constant at FWHM=1\,\kms\ (equal to thermal broadening at $\sim$1000\,K) and FWHM=6\,\kms, respectively.  

\begin{figure}
\includegraphics[angle=270,width=\columnwidth]{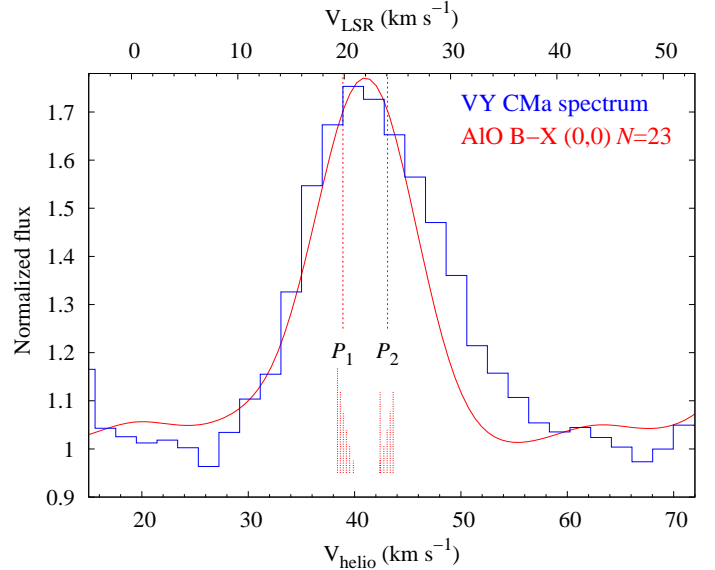}
\caption{The $P$(23) rotational component of the $B$--$X$ (0,0) band of AlO in the simulation (red line) and in the observed spectrum of VY\,CMa (blue histogram). The velocity is given with respect to the average wavelength of the doublet, whose splitting is 4.2\,\kms. The positions of the doublet components are shown by the two dashed lines in the upper part of the figure. Lines in the lower part show the hyperfine splitting in both components \citep[from][]{hyper}. }
\label{fig-AlO-one-rotational-profile}
\end{figure}

The vibrational temperature ($T_{\rm vib}$) is indicated by the relative strengths of heads within each sequence. We found satisfactory results for a broad temperature range between 1600\,K and 2800\,K. Particularly good fits to data were obtained when different vibrational temperatures (but still within the specified range) were assumed for different bands. The best simulation with $T_{\rm rot}$=700\,K and $T_{\rm vib}$=2200\,K is shown in Figs.\,\ref{fig-AlO-bands} and \ref{fig-AlO-one-rotational-profile}.

By comparing the simulation to observations, we were able to derive the bulk velocity of the AlO gas. Using the cross-correlation method implemented in IRAF's task {\it fxcor}, we derived heliocentric radial velocity of $V_{\rm helio}$=41$\pm$1\,\kms\ (or equivalently $V_{\rm LSR}$=21$\pm$1\,\kms\ with respect of the Local Standard of Rest). This velocity is confirmed when peaks of individual rotational components are compared to their rest wavelengths. Moreover, we do not see any systematic shift in peak positions with increasing excitation energy of the rotational components. 

\begin{figure}
\includegraphics[angle=270,width=\columnwidth]{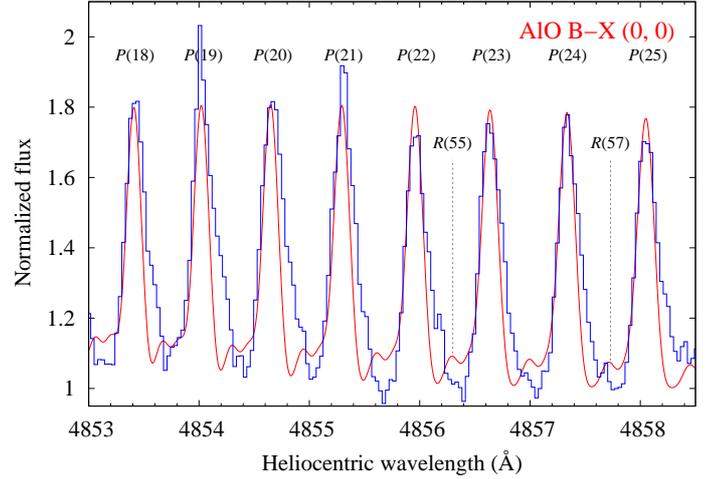}
\caption{Rotational components of the $B$--$X$ (0,0) band of AlO as observed in VY\,CMa (blue histogram) and in our simulation (red line). The individual lines are unresolved doublets of members of the $P_1$ and $P_2$ branches. Their peaks are designated by rotational quantum numbers. Much fainter rotational components of the $R$ branch can be seen in the simulation (only two were marked in the figure).}
\label{fig-AlO-rotational-lines}
\end{figure}

Because of the good separation of the rotational components and their low optical thickness at high $N$ (usually $\tau\!\ll$1), we can examine the shape of individual lines for the kinematical profile of the AlO-bearing gas. The measured FWHM of lines between $P$(20) and $P$(30) of the (0,0) band is 14.4$\pm$1.2\,\kms. This value is however affected by several splitting effects. The nuclear hyperfine splitting due to nuclear spin of $I$=5/2, which was not included in the simulation, is known to be very small \citep[about 1.5\,\kms\ at $P$(23);][]{hyper}, i.e. smaller than our spectral resolution. Additionally, each rotational line is a blend of two components belonging the $P_1$ and $P_2$ branches separated due to the $\rho$-splitting, which is very well known for the (0,0) band. Its magnitude changes linearly with rotational number and is negligible at low $N$ (e.g. 0.3\,\kms\ at $N$=2), but significant at high rotational numbers (e.g. 6\,\kms\ at $N$=33). Correcting the measured widths for $\rho$-doubling gives a typical FWHM of 13.5$\pm$1.5\,\kms. To illustrate the different splitting effects on the line profile we show them graphically in Fig.\,\ref{fig-AlO-one-rotational-profile} for the $P$(23) line for which the hyperfine structure is known from \citet{hyper}. 

A comparison of the individual profiles to the simulation in Fig.\,\ref{fig-AlO-rotational-lines} reveals a small asymmetry of the observed profiles, i.e. the red wing is more extended (by $\sim$2\,\kms). Although there is some uncertainty in the molecular data which we used in the simulation, we believe this effect is real. Additionally, on the blue side of each rotational component, a very weak absorption can be seen (it is more apparent when the spectrum is compared to the simulation -- see Figs.\,\ref{fig-AlO-rotational-lines}--\ref{fig-AlO-one-rotational-profile}). This and the profile asymmetry suggest that each rotational component may have a P-Cyg type profile. Some absorption can be also seen next to each of the strong band-heads (Fig.\,\ref{fig-AlO-bands}), what supports this interpretation. The absorption is however very weak and the P-Cyg shape is claimed here only as a tentative result.

\section{Discussion}  \label{dis}
\subsection{Interpretation of the emission properties} \label{dis1}
The central velocity of the optical AlO emission ($V_{\rm LSR}$=21$\pm$1\,\kms) is consistent with the velocity of the AlO millimetre lines (25$\pm$6\,\kms) and is in excellent agreement with the stellar systemic velocity known from optical and radio/(sub-)millimetre observations ($V_{\rm LSR}$=21.0$\pm$1.5\,\kms) \citep[e.g.][and references therein]{smithKI}. Although, the central velocity of the AlO emission is much more accurately measured in the optical (where splitting effects are smaller and a large number of rotational transitions at high signal-to-noise ratio is observed simultaneously), the centres of optical emission lines may be red-shifted due to scattering on dust grains in the thick and expanding nebula \citep{vanBlerkom,romanik}; this effect might have been observed in some absorption and emission features from the outer nebula of VY\,CMa \citep{h05}. However, since the position of radio and optical lines of AlO are consistent within the uncertainties, this scattering effect must be very small in our case ($\lesssim$3\,\kms), i.e. certainly much smaller than suggested theoretically by \citet{vanBlerkom} for photospheric absorption lines in VY\,CMa. 
 
The outflow of VY\,CMa observed in millimetre emission lines has been ascribed to at least three distinct kinematical components \citep{tanen10} with the fastest material moving at expansion velocity of at least 40\,\kms\ producing broad lines with multiple peaks. In addition to those wide features, narrow emission lines centered approximately at the stellar systemic velocity have been detected, with the narrowest emission known arising from NaCl with FWHM=16$\pm$3\,\kms\ \citep{detect_NaCl}; TiO is another molecule which displays such narrow features with FWHM=18$\pm$2\,\kms\ \citep{kami_tio}. The  narrowest of the pure atomic emission lines present in UVES spectrum have a FWHM of 22\,\kms. All these narrow lines are signatures of gas located within the inner outflow, perhaps even below the dust formation zone. The width of the AlO emission of 13.5\,\kms\ is smallest among all detected (thermal) features in VY\,CMa's circumstellar environment indicating that AlO probes deepest observed layers of the material leaving the photosphere. With the typical full-width at zero intensity of 27$\pm$2\,\kms\ corrected for the $\rho$-doubling, the AlO emission implies gas expansion velocity of $V_{\rm exp}$=13.5$\pm$2.0\,\kms.

It should be mentioned that maser lines of SiO (and to a lesser degree of H$_2$O) are known to exhibit even narrower emission components with widths down to 1\,\kms. These maser lines undoubtedly arise in the inner, accelerating envelope of VY\,CMa (within tens of mas from the star), but they usually do not trace the whole outflow, as most of thermal lines do, but rather individual clouds within the outflow. Moreover, due to a complex excitation mechanism, the SiO masers are observed in a ring-like feature surrounding the photosphere of VY\,CMa \citep[e.g.][]{zhang} showing complex dynamics. Hence, the half-widths of individual lines do not correspond directly to the expansion velocity. 

The slight asymmetry observed in AlO line profiles, with the red wing being more extended, is reminiscent of what is observed in the (sub-)millimetre lines of TiO \citep{kami_tio}. This asymmetry may suggest that even the inner wind region exhibits some degree of asymmetry. However, exactly the same type of asymmetry is expected to arise due to scattering in an expanding dusty nebula \citep{vanBlerkom,romanik}. In addition to the optical thickness and geometry of the scattering medium, this effect is strongly dependent on the scattering properties of dust  and for some dust types produces profile asymmetry without any apparent shift in line centres \citep{romanik}, which could be the case for the AlO lines. The origin of the asymmetry is not clear and requires modelling including a detailed treatment of scattering on dust. This is beyond the scope of this work. We note, however, that if the observed profiles are re-shaped by scattering, they are also slightly broadened and the intrinsic width of the AlO emission is even smaller than quoted above. The pure rotational lines of AlO observed at millimetre wavelengths by \citet{tanen_AlO}, for which scattering can be completely neglected,  are severely affected by the hyperfine splitting and modest signal-to-noise ratios which complicate the profile analysis and the asymmetry cannot be verified. However, the authors were able to constrain the line  widths to 10--15\,\kms\ what is consistent with our analysis of the optical profiles. 
  
The vibrational temperature we derived describes the population of vibrational levels in the upper electronic state in the sense of the Boltzmann distribution. Because the excitation is most likely radiative (the electronic state $B^2\Sigma$ is 29\,691\,K above the ground level), this approach is a crude simplification. The population of vibrational levels in $B^2\Sigma$ depends on the population in the ground state, the spectrum of the external radiation field, and the strengths of absorption bands populating the vibration levels. By taking Franck-Condon factors as band intensities and assuming a flat stellar spectrum, we calculated that a temperature between 1200 and 1800\,K is more representative as the vibrational temperature. Taking the large uncertainties and the devious meaning of $T_{\rm vib}$, it is of no practical value for the further discussion.

On the other hand, the rotational temperature of 700\,K is expected to be  close to the gas kinetic temperature, $T_{\rm kin}$, what makes it a valuable parameter. The rotational temperature we derive is much above that found by \citet{tanen_AlO} ($T_{\rm rot}\!\sim$250\,K) on the basis of a rotational-diagram analysis. Their value is however very uncertain as only three lines were measured at low signal-to-noise ratio. Our value of $T_{\rm rot}$ is consistent with temperatures derived from analyses of other optical bands of refractory molecules observed in VY\,CMa in emission --- \cite{phillips} found $T_{\rm rot}$=600$\pm$150\,K from an analysis of the TiO $\gamma^{\prime}$ system in emission, while \cite{herbig74} found $T_{\rm rot}$=370 and 790\,K  for the orange ($A$--$X$) band of ScO observed at two different dates. 

\subsection{The origin of the AlO emission}
In the 2-dimensional UVES spectrum, the AlO emission is unresolved meaning that the emission region is smaller than 0\farcs8 FWHM. The width of the AlO profile allows to constrain even better the size of the emitting region. By combining tangential (proper) and radial motions observed in H$_2$O masers at three epochs, \citet{richards} found expansion of the envelope within 75 to 440\,mas from the star. By applying their expansion model (updated to the stellar distance of 1.2\,kpc) we get a radius $r$=18$\pm$6\,R$_{\star}$ (R$_{\star}$=2000\,R$_{\sun}$) for the AlO-emitting region. 

The type of observed line profiles also provides strong constraints on the dimension of the AlO emission region. If the gas was much closer to the star than the 18\,R$_{\star}$ found above, say a few $R_{\star}$, the optical rotational profiles should display pronounced absorption components arising in gas seen in front of the stellar disk. Combined with a velocity gradient (which is observed in this region in SiO maser lines), such configuration should produce P-Cyg profiles with a high absorption-to-emission intensity ratio, i.e. much more pronounced than those we observe (if any absorption is present at all).  

\citet{decin} modelled the envelope of VY\,CMa with a variable mass loss and derived thermal model of the envelope with a detailed treatment of heating and cooling mechanisms. Assuming that $T_{\rm kin} \! \approx \! T_{\rm rot}$, we find that the temperature of 700$\pm$200\,K  corresponds to a distance of 20$\pm$5\,R$_{\star}$ in \citet{decin}, which is consistent with the constraints found above. In the same model, gas   of such temperature is located in the accelerating outflow. Their velocity law applied to the AlO expansion velocity ($V_{\rm exp}$=13.5\,\kms) indicates location within $\sim$10\,R$_{\star}$. The hydrogen density at 10--25\,R$_{\star}$ in that model (corresponding to the mass-loss rate of 10$^{-4}$\,M$_{\sun}$\,yr$^{-1}$) is within $n_{\rm H}$=10$^7$--10$^9$\,cm$^{-3}$. Because critical densities for the observed rotational transitions of AlO are between 5$\cdot$10$^5$\,cm$^{-3}$ (for $N\!\!\approx\!\!1$) and 5$\cdot$10$^9$\,cm$^{-3}$ (for $N\!\!\approx\!\!20$)\footnote{The critical densities for the AlO rotational transitions in the $X^2\Sigma$ state were calculated for the collisional cross-section $\sigma$=10$^{-15}$\,cm$^2$ and gas thermal broadening of 1\,\kms.}, the population of rotational levels has a chance to be thermalized by collisions in this region \citep[cf.][]{woitke99}, justifying our assumption that $T_{\rm kin} \! \approx \! T_{\rm rot}$. We must however note that the model of \citet{decin} should be treated with caution as its parameters were constrained by observations of CO lines with the highest transition being $J$=7--6 (155\,K above the ground, $n_{\rm crit}$=4$\cdot$10$^5$\,cm$^{-1}$) which does not probe well the accelerating  outflow.  

The above discussion suggests that the AlO emission comes from a region of a diameter of about 40\,R$_{\star}$ (or 0\farcs4). This allows us to calculate the AlO mean abundance. From our simulations, we get the AlO total column density of $N_{\rm AlO}$=6$\cdot$10$^{16}$\,cm$^{-2}$. This value has a factor of a few uncertainty resulting from the uncertainties in the other simulation parameters. For the mass-loss rate of 10$^{-4}$\,M$_{\sun}$\,yr$^{-1}$, the envelope extending from 1\,R$_{\star}$ to 20\,R$_{\star}$, the mean molecular weight of 1.3, and the mean velocity of 6\,\kms, the total hydrogen column is $N_{\rm H}$=3$\cdot$10$^{24}$\,cm$^{-2}$ giving AlO mean abundance of [$f$]=$\log(N_{\rm AlO}/N_{\rm H})\!\approx$--7.7. This value is uncertain by at least 1\,dex. 

Chemical equilibrium models of circumstellar gas around evolved oxygen-rich stars predict high abundances of AlO only at relatively high temperatures. For instance, in the chemical-equilibrium model calculated by \citet{inorganicdust}, the AlO abundance is expected to be higher than $[f]$=--9 only at temperatures between 1100 and 2000\,K. Below these temperatures, the abundance of AlO drops steeply with the distance from the star (being converted to Al$_2$O, Al$_2$O$_2$, and Al$_2$O$_3$) and at 800\,K is already 6\,dex below its peak value ([$f_{\rm max}]\!\approx$--8.5). Including condensation of alumina species into such models results in even stronger depletion of AlO at temperatures lower than 1800\,K \citep{SaH}. According to these models, therefore, the AlO gas should be located in a relatively thin layer a few $R_{\star}$ above the photosphere. This is strongly contradicted by our constraints on the size of the AlO emission region \citep[see also][]{tanen_AlO} suggesting that the heavier alumina compounds, e.g. Al$_2$O$_3$, are formed less effectively than the models predict. Because Al$_2$O$_3$ is expected to condense to corundum, which is often thought to be the first circumstellar condensate \citep{SaH,l03}, this disagreement may signalize that we should revise the picture of how effectively (if at all) alumina dust is formed in oxygen-rich stars like VY\,CMa.  

That the chemical-equilibrium models fail to explain the aluminium chemistry in the envelope of VY\,CMa has been already noted in \citet{tanen_AlO}. They proposed that the discrepancy may be caused by ``freezing out" of the equilibrium abundances at a few R$_{\star}$ or that shocks disrupt the process of molecule formation at larger distances. ``Freezing out" \citep{freezing-out} occurs when the time scale of chemical reactions ($\tau_{\rm chem}\!\propto\!(Kn)^{-1}$, where $K$ is the reaction rate) becomes much longer than the dynamical scale of the outflow ($\tau_{\rm dyn}\!\approx\!r/V_{\rm exp}$). As long as chemical reactions are faster than the dynamical time scale, the molecular abundances are thought to be well described by equilibrium values, but -- as the density drops -- at some point in the outflow the reactions become inefficient and abundances do not change any more being fixed at a certain level corresponding to some equilibrium values. Following  \citet{freezing-out}, we estimate that such freezing-out should occur in the outflow of VY\,CMa at densities of about $n_{\rm H}$=10$^{10}$\,cm$^{-3}$, which are expected at a distance of few R$_{\star}$ \citep{decin}. Such a ``freezing out" scenario in VY\,CMa would be supported for AlO by its abundance being close to the maximal value predicted by equilibrium models (and indeed those maximal abundances correspond to distances of a few R$_{\star}$). Whether such process can really take place in VY\,CMa should be verified by constructing a detailed model which would take into account the variable conditions in the outflow. 

We favour the alternative scenario in which the observed over-abundance of AlO at larger radii is a result of shocks related to the stellar activity. There are strong suggestions that shocks are present in the close vicinity of VY\,CMa:
\begin{itemize}
\item It has been proposed that the complex nebula surrounding VY\,CMa was produced by a violent and episodic process that most likely involves creation of shocks (\citealp[e.g. episodic mass-ejections advocated in][]{smith01} and \citealp{h05,humphreys07}). This proposal, however,  is based on observations of the extended nebula, formed by at least 1000\,yr of mass-loss history, and it is not sure whether the same kind of activity is currently taking place near the stellar surface.
\item The presence of extended and variable emission and broad absorption of H$\alpha$ \citep[e.g.][]{h05} within the otherwise neutral nebula indicates the presence of a process energetic enough to ionise hydrogen (although some of the emission, especially in the outermost nebula, may arise in the nearby \ion{H}{II} region).
\item The star is known as an irregular variable, with changes of the optical light reaching 2\,mag. This variability may be partially explained  by (non-radial) pulsations \citep{h05}, which would be a potential driving source of shocks near the stellar surface. 
\item As an extreme supergiant very close to its Hayashi limit \citep{wittkowski}, VY\,CMa has huge convective cells \citep{schwarzschild}, which are expected to severely affect its circumstellar medium \citep{smith01,h05,humphreys07}. 
\item {\it Circumstellar} AlO bands are rarely observed. Their presence and erratic behaviour has been long known in Mira variables \citep{keenan} and was reported in an enigmatic object U\,Equ \citep{uequ}. They have been also observed in some members of the recently recognized family of nova-like variables called `red novae'. These include V4332\,Sgr, which has the optical and NIR bands of AlO in emission \citep{banerjee03,tyl_v4332}, and V838\,Mon, which shows  circumstellar absorption in those bands \citep{kami_v838_keck,tylenda_uves,evans,lynch}. The AlO emitters seem to share one common characteristics -- their circumstellar environments have been shocked by violent motions, e.g. large-amplitude pulsations in case of Miras, and explosions in case of red novae and U\,Equ. The presence of shocks would justify VY\,CMa belonging to the group of AlO-emitters. 
\end{itemize}
It would be indeed very surprising if the star were not to develop shocks near its surface. As shown by \citet{cherchneff} (for asymptotic giant branch stars), propagation of shocks through material above the stellar photosphere changes its chemical content considerably with respect to what is expected in chemical equilibrium. Future models attempting to explain the chemistry in VY\,CMa's envelope should explore this non-equilibrium approach. Since shocks are expected to produce transient phenomena, there is also a possibility of observational verification. Most of the previously described molecular bands seen in VY\,CMa in emission are variable \citep[e.g.][]{waller86,herbig74}, although it is not clear whether the absolute intensity or the intensity relative to the variable continuum changes. It is also not known what are the shortest time scales on which the variability occurs except that it may be shorter than years. To explore the possibility that these optical emission bands arise due to the presence of shocks, it is highly desirable to verify whether the AlO bands are variable too and whether there is any correlation with the variability of the other molecular bands and hydrogen recombination lines.

\acknowledgements{We are very grateful to G. Wallerstein and G. Gonzalez for making available to us their optical spectrum of VY\,CMa. This work makes use of data acquired within the ESO programmes 67.B-0504 and DDT 266.D-5655, and observations from the AAVSO International Database. The research was supported by the PL-Grid infrastructure and partially founded from the National Science Center grant DEC-2011/01/B/ST9/02229.}


\begin{thebibliography}{}
\bibitem[Banerjee et al.(2003)]{banerjee03} 
Banerjee, D.~P.~K., Varricatt, W.~P., Ashok, N.~M., \& Launila, O.\ 2003, \apjl, 598, L31 


\bibitem[Bagnulo  et al.(2003)]{POP} 
Bagnulo, S., Jehin, E., Ledoux, C., et al.\ 2003, The Messenger, 114, 10

\bibitem[Barnbaum et al.(1996)]{uequ} 
Barnbaum, C., Omont, A., \& Morris, M.\ 1996, \aap, 310, 259 

\bibitem[Buhl et al.(1974)]{SiOdetect} 
Buhl, D., Snyder, L.~E., Lovas, F.~J., \& Johnson, D.~R.\ 1974, \apjl, 192, L97 

\bibitem[Cherchneff(2006)]{cherchneff} 
Cherchneff, I.\ 2006, \aap, 456, 1001 

\bibitem[Coxon \& Naxakis(1985)]{coxon} 
Coxon, J.~A., \& Naxakis, S.\ 1985, J. Mol. Spec., 111, 102 

\bibitem[Danchi et al.(1994)]{danchi} 
Danchi, W.~C., Bester, M., Degiacomi, C.~G., Greenhill, L.~J., \& Townes, C.~H.\ 1994, \aj, 107, 1469 

\bibitem[Decin et al.(2006)]{decin} 
Decin, L., Hony, S., de Koter, A., Justtanont, K., Tielens, A.~G.~G.~M., \& Waters, L.~B.~F.~M. \ 2006, \aap, 456, 549 

\bibitem[Eliasson \& Bartlett(1969)]{OHdetect} 
Eliasson, B., \& Bartlett, J.~F.\ 1969, \apjl, 155, L79 

\bibitem[Evans et al.(2003)]{evans} 
Evans, A., Geballe, T.~R., Rushton, M.~T., et al.\ 2003, \mnras, 343, 1054 

\bibitem[Gail \& Sedlmayr(1998)]{inorganicdust} 
Gail, H.-P., \& Sedlmayr, E.\ 1998, Faraday Discussions, 109, 303 


\bibitem[Herbig(1970)]{herbig1970} Herbig, G.~H.\ 1970, Memoires 
of the Societe Royale des Sciences de Liege, 19, 13 

\bibitem[Herbig(1974)]{herbig74} 
Herbig, G.~H.\ 1974, \apj, 188, 533 

\bibitem[Hougen(1970)]{hougen} 
Hougen, J.T. 1970, The Calculation of Rotational Energy Levels and Rotational Line Intensities in Diatomic Molecules, Nat. Bur. Stand. (U.S.),  Monogr. 115 (June 1970)

\bibitem[Humphreys et al.(2005)]{h05} 
Humphreys, R.~M., Davidson, K., Ruch, G., \& Wallerstein, G.\ 2005, \aj, 129, 492 

\bibitem[Humphreys et al.(2007)]{humphreys07} 
Humphreys, R.~M., Helton, L.~A., \& Jones, T.~J.\ 2007, \aj, 133, 2716 

\bibitem[Hyland et al.(1969)]{hyland69} 
Hyland, A.~R., Becklin, E.~E., Neugebauer, G., \& Wallerstein, G.\ 1969, \apj, 158, 619 


\bibitem[Joy(1942)]{joy42} 
Joy, A.~H.\ 1942, \apj, 96, 344 

\bibitem[Kami{\'n}ski et al.(2009)]{kami_v838_keck} 
Kami{\'n}ski, T., Schmidt, M., Tylenda, R., Konacki, M., \& Gromadzki, M.\ 2009, \apjs, 182, 33 

\bibitem[Kami{\'n}ski et al.(2012)]{kami_tio} 
Kami{\'n}ski, T. et al.\ 2012, \aap\ submitted 

\bibitem[Kami{\'n}ski et al.(2010)]{kami_v4332} 
Kami{\'n}ski, T., Schmidt, M., \& Tylenda, R.\ 2010, \aap, 522, A75 

\bibitem[Keenan et al.(1969)]{keenan} 
Keenan, P.~C., Deutsch, A.~J., \& Garrison, R.~F.\ 1969, \apj, 158, 261 

\bibitem[Knowles et al.(1969)]{H2Odetect} 
Knowles, S.~H., Mayer, C.~H., Sullivan, W.~T., III, \& Cheung, A.~C.\ 1969, Science, 166, 221 

\bibitem[Kovacs(1969)]{kovacs} 
Kovacs, I.\ 1969, Rotational structure in the spectra of diatomic molecules, London (UK): Adam Hilger  

\bibitem[Launila \& Berg(2011)]{hyper} 
Launila, O., \& Berg, L.-E.\ 2011, J. Mol. Spec. 265, 10 

\bibitem[Lodders(2003)]{l03} 
Lodders, K.\ 2003, \apj, 591, 1220 

\bibitem[Lynch et al.(2004)]{lynch} 
Lynch, D.~K., Rudy, R.~J., Russell, R.~W., et al.\ 2004, \apj, 607, 460 

\bibitem[Massey et al.(2006)]{massey} 
Massey, P., Levesque, E.~M., \& Plez, B.\ 2006, \apj, 646, 1203 

\bibitem[McCabe et al.(1979)]{freezing-out} 
McCabe, E.~M., Smith, R.~C., \& Clegg, R.~E.~S.\ 1979, \nat, 281, 263 

\bibitem[Milam et al.(2007)]{detect_NaCl} Milam, S.~N., Apponi, 
A.~J., Woolf, N.~J., \& Ziurys, L.~M.\ 2007, \apjl, 668, L131 


\bibitem[Phillips \& Davis(1987)]{phillips} 
Phillips, J.~G., \& Davis, S.~P.\ 1987, \pasp, 99, 839 


\bibitem[Richards et al.(1998)]{richards} 
Richards, A.~M.~S., Yates, J.~A., \& Cohen, R.~J.\ 1998, \mnras, 299, 319 

\bibitem[Romanik \& Leung(1981)]{romanik} 
Romanik, C.~J., \& Leung, C.~M.\ 1981, \apj, 246, 935 

\bibitem[Royer et al.(2010)]{royer} 
Royer, P., Decin, L., Wesson, R., et al.\ 2010, \aap, 518, L145 

\bibitem[Saksena et al.(2008)]{saksena} 
Saksena, M.~D., Deo, M.~N., Sunada, K., Behere, S.~H., Londhe, C.~T. 2008, J. Mol. Spec., 247, 47

\bibitem[Schwarzschild(1975)]{schwarzschild} 
Schwarzschild, M.\ 1975, \apj, 195, 137 

\bibitem[Sharp \& Huebner(1990)]{SaH} 
Sharp, C.~M., \& Huebner, W.~F.\ 1990, \apjs, 72, 417 

\bibitem[Smith et al.(2001)]{smith01} 
Smith, N., Humphreys, R.~M., Davidson, K., et al.\ 2001, \aj, 121, 1111 

\bibitem[Smith(2004)]{smithKI} 
Smith, N.\ 2004, \mnras, 349, L31 


\bibitem[Tenenbaum \& Ziurys(2009)]{tanen_AlO} 
Tenenbaum, E.~D., \& Ziurys, L.~M.\ 2009, \apjl, 694, L59 

\bibitem[Tenenbaum et al.(2010)]{tanen10} 
Tenenbaum, E.~D., Dodd, J.~L., Milam, S.~N., Woolf, N.~J., \& Ziurys, L.~M.\ 2010, \apjs, 190, 348 

\bibitem[Tylenda et al.(2005)]{tyl_v4332} 
Tylenda, R., Crause, L.~A., G{\'o}rny, S.~K., \& Schmidt, M.~R.\ 2005, \aap, 439, 651 


\bibitem[Tylenda et al.(2011)]{tylenda_uves} 
Tylenda, R., Kami{\'n}ski, T., Schmidt, M., Kurtev, R., \& Tomov, T.\ 2011, \aap, 532, A138 

\bibitem[van Blerkom \& van Blerkom(1978)]{vanBlerkom} 
van Blerkom, J., \& van Blerkom, D.\ 1978, \apj, 225, 482 

\bibitem[Wallerstein(1971)]{waller71} 
Wallerstein, G.\ 1971, \apj, 169, 195 

\bibitem[Wallerstein(1986)]{waller86} 
Wallerstein, G.\ 1986, \aap, 164, 101 

\bibitem[Wallerstein \& Gonzalez(2001)]{waller2001}
Wallerstein, G., \& Gonzalez, G.\ 2001, \pasp, 113, 954 

\bibitem[Wittkowski et al.(2012)]{wittkowski} 
Wittkowski, M., Hauschildt, P.~H., Arroyo-Torres, B., \& Marcaide, J.~M.\ 2012, \aap, 540, L12 

\bibitem[Woitke et al.(1999)]{woitke99} 
Woitke, P., Helling, C., Winters, J.~M., \& Jeong, K.~S.\ 1999, \aap, 348, L17 

\bibitem[Zenouda et al.(1999)]{zeno} 
Zenouda, C., Blottiau, P., Chambaud, G., Rosmus, P.\ 1999, J. Mol. Struct.-Theochem., 458, 61

\bibitem[Zhang et al.(2012)]{zhang} 
Zhang, B., Reid, M.~J., Menten, K.~M., \& Zheng, X.~W.\ 2012, \apj, 744, 23 

\bibitem[Ziurys et al.(2007)]{ziurys} 
Ziurys, L.~M., Milam, S.~N., Apponi, A.~J., \& Woolf, N.~J.\ 2007, \nat, 447, 1094 

\end{thebibliography}
\end{document}